# Viscoelastic peeling of thin tapes with frictional sliding

Marco Ceglie[1][0000-0002-6844-6768] , Nicola Menga[1][0000-0002-4728-1773]
and Giuseppe Carbone[1][0000-0002-8919-6796]

[1] Department of Mechanics, Mathematics and Management, Polytechnic University of Bari, Italy
`marco.ceglie@poliba.it`

**Abstract.** Peeling is one of the most common detachment mechanisms adopted in industrial applications. However, although several experimental investigations have proven the possible occurrence of relative sliding at the interface close to the peeling front, a comprehensive model considering the effect of both the tape viscoelasticity and frictional interfacial dissipation on the peeling behavior is lacking. The present study aims at providing a theoretical framework to investigate the peeling process of a thin viscoelastic tape from a rigid substrate in the presence of frictional sliding at the interface. It shows that, under certain conditions, significantly tougher adhesive performance can be achieved compared to stuck elastic conditions with no interfacial sliding, and that the delamination resistance of the system strongly depends on the propagation velocity.

**Keywords:** Peeling, Viscoelasticity, Interfacial friction, Adhesion.

## 1. Introduction

Peeling is one of the most exploited mechanism of detachment both in nature and human-made context. It plays a crucial role in essential functions and locomotion of various animals, as well as in regulating adhesion within biological systems [1-3]. Furthermore, this mechanism finds significance in technological domains, spanning applications in areas like painting and coating [4], transfer printing [5], and the development of soft grippers in robotics [6]. The relevance of peeling is further underscored by its application in material characterization and measuring adhesive forces through peeling tests [7].

When the peeling process is employed to explain the adhesive strength of real systems, most of the theories rely on the Kendall' model derived for pure elastic tapes peeled away from rigid substrates with the interfaces in completely stuck contact [8]. Later on, this theory has been extended to the case V peeling of elastic tapes from soft substrate of infinite [9] or finite thickness [10]. However, following evidence of sliding between tape and substrate near the crack front during peeling propagation [11], many studies have tried to consider this behavior relying on mode-mixity approaches [12] or energy based formulation which extends Kendall's original model by accounting for frictional dissipation energy contribution. In the latter scenario, this problem has been addressed within the context of linear elasticity considering both small and large deformations, as well as single and V-peeling configurations [13-15]. Theoretical and experimental investigations have consistently concluded that the presence of frictional sliding at the interface enhances the peeling resistance of the system, resulting in higher peeling forces. Nevertheless, while this dissipation mechanism appears to be pivotal in governing peeling phenomena, it is often ignored, and it has not been comprehensively explored. Similarly, as in several other tribological phenomena [16], viscoelasticity also plays a key role, leading to unsteady V-peeling behavior [17].

In this study, building on the formalism presented in Ref. [18], we investigate the frictional peeling of a viscoelastic tape from a rigid substrate, aiming at highlighting the interplay between frictional sliding, bulk viscous dissipation in governing the peeling behavior.

# 1 Formulation

The peeling model for a thin viscoelastic tape of thickness $d$ and width $b$ (cross ection $A_t$) from a rigid substrate is here presented (see Fig. 1). The tape detachment is driven by a peeling force $P$ applied at a constant peeling angle $\theta$. We consider a steady-state process with a constant propagation velocity $v_c$. For simplicity, a linear viscoelastic material with a single relaxation time $\tau$ is assumed, having $E_0$ and $E_\infty$ as low and high frequency moduli, respectively. The crack propagation condition is derived from an energy balance. Dynamic effect and minor dissipative contributions will be neglected (such as acoustic and thermal emissions). Two different conditions are considered: (i) stuck and (ii) sliding interface.

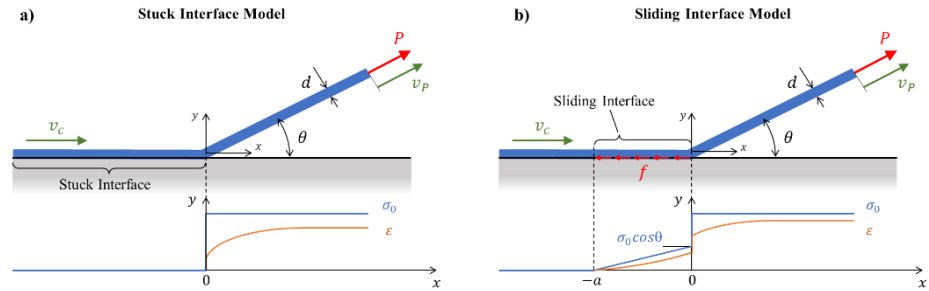

**Fig. 1.** Schemes of stuck interface (a) and sliding interface (b) models for the peeling process of a thin viscoelastic tape from a rigid substrate. Qualitative diagrams of the tape stress $\sigma$ (blue) and deformation $\varepsilon$ (orange) are shown in the lower part.

## 1.1 Stuck interface

In this scenario, the tape firmly sticks to the substrate so that no tangential slidings occur at the interface during the delamination process, and the adhering region of the tape results unloaded (Fig. 1a). The system energy balance per unit time is given by

$$W_P + W_{in} + W_{ad} = 0 \qquad (1)$$

where $W_P$ is the work per unit time of the external forces acting on the tape, $W_{in}$ is the work per unit time done by tape internal stresses, and $W_{ad}$ is the work per unite time associated with the adhesive forces.

Referring to a reference frame centered in the peeling front, the external forces contribution accounts for the work of the peeling force $P$, and the substrate tangential reaction $-P\cos\theta$, so that

$$W_P = Pv_p - Pv_c \cos\theta = \sigma_0 v_c A_t (1 + \varepsilon_P - \cos\theta) \qquad (2)$$

being $\sigma_0 = P/A_t$ the tape stress, and $\varepsilon_P = \sigma_0/E_0$ the tape strain at the tip assumed completely relaxed. Note that, for the tape mass balance, $v_p = v_c(1 + \varepsilon_P)$.

The internal energy contribution accounts for the elastic energy stored in the detached tape and the losses due to the viscoelastic creep. The change of stress across the peeling front is modeled as a step-function, so that $\sigma(x) = H(x)\sigma_0$, whereas the strain field can be derived for a steady-propagation as $\varepsilon(x) = \int_{-\infty}^{x} J(x-s)\frac{d\sigma}{ds}(s)ds$ [19], $J(x) = E_0^{-1} - (E_\infty^{-1} - E_0^{-1})e^{-x/v_c\tau}$ being the creep function. Focusing on tensile contribution, we have

$$W_{in} = -v_c A_t \int_{-\infty}^{\infty} \sigma(x)\frac{d\varepsilon}{dx}(x)dx = -v_c A_t \sigma_0^2 \frac{2E_\infty - E_0}{2E_\infty E_0} \qquad (3)$$

The surface term $W_{ad}$ represents the energy per unit time associated with the rupture of interfacial bonds between the tape and the substrate; being $\gamma$ the adhesion energy, we have

$$W_{ad} = -v_c w\, \gamma \qquad (4)$$

Combining Eqs. (1,2,3,4), the peeling propagation condition is given by

$$\frac{P^2}{2\,E_\infty w^2\, d} + \frac{P}{w}(1 - \cos\theta) = \gamma \qquad (5)$$

which resembles the Kendall equation [8] with the elastic modulus given by the high frequency viscoelastic modulus $E_\infty$. Note that it does not depend on the propagation velocity.

## 1.2 Sliding interface

In fact, in real peeling process relative slidings between tape and substrate occur due to tape deformation, leading to interfacial shear stresses which balance the tangential component of the peeling force (Fig. 1b). Assuming a uniform shear stress value $f$, the force equilibrium gives the extension of the sliding region $a = P \cos \theta / w f$, and the stress field within the tape: $\sigma(x) = \frac{f}{d}(x + a)$ for $-a < x < 0$, and $\sigma(x) = \sigma_0$ for $x > 0$, which allow to evaluate the strain field $\varepsilon(x)$ through the constitutive relation already used above.

In this case, the energy balance is given by

$$W_P + W_{in} + W_{ad} + W_T = 0 \qquad (6)$$

which accounts for an additional internal energy contribution associated with the tape deformation in the sliding zone (evaluated as in the stuck interface case), and for the frictional energy dissipation per unit time given by

$$W_T = -\int_{-a}^{0} v_s(x) f \, dx = -f w v_c \int_{-a}^{0} \varepsilon(x) dx \qquad (7)$$

Finally, the propagation condition results

$$\frac{P^2}{2E_0 w^2 d}\left\{(1 - \cos^2 \theta) - \frac{\kappa - 1}{\kappa}(1 - \cos \theta)\left(1 + 2 \cos \theta \left[\frac{v_c \tau}{a}\left(1 - e^{-\frac{a}{v_c \tau}}\right) - \frac{1}{2}\right]\right)\right\} + \frac{P}{w}(1 - \cos \theta) = \gamma \qquad (8)$$

which clearly depends on the peeling velocity, the frictional features of the interface and material parameters.

## 2 Results

In order to simplify the interpretation of the results, we refer to the following dimensionless parameters: $\tilde{P} = P/A_t E_0$, $\tilde{f} = f/E_0$, $\tilde{\gamma} = \gamma/E_0 d$ and $\tilde{v}_c = v_c \tau/d$. In our calculations reasonable values of dimensionless parameters are assumed consistently with experimental results taken from literature [20-22].

## 2.1 Effect of viscoelastic parameter and adhesion energy

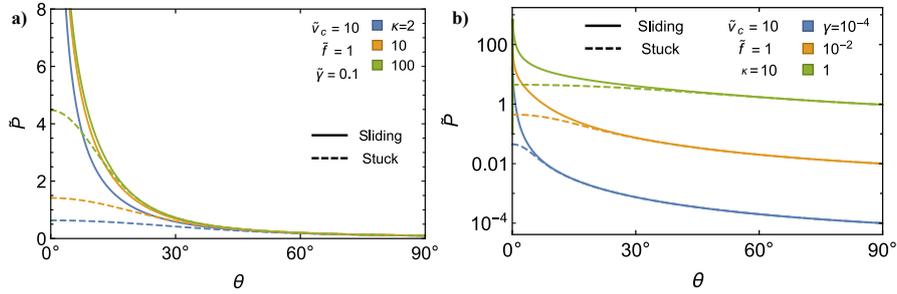

**Fig. 2.** The dimensionless peeling force $\tilde{P}$ as a function of the peeling angle $\theta$, for different values of (a) the viscoelastic parameter $\kappa = E_\infty/E_0$, and (b) the dimensionless work of adhesion $\tilde{\gamma}$ predicted by stuck (dashed line) and siding (solid line) interface.

   The overall peeling behavior predicted by the stuck and sliding models is depicted in Fig. 2, where the dimensionless peeling force $\tilde{P}$ is given as a function of the viscoelastic parameter $\kappa = E_\infty/E_0$ and the dimensionless adhesion energy $\tilde{\gamma}$. As already noted, in the case of stuck interface, a Kendall like behavior is recovered with the peeling force being a decreasing function of the peeling angle. However, if a finite peeling force $P \to w\sqrt{2\gamma E_\infty}$ is predicted for a vanishing value of the peeling angle by the stuck model, the presence of interfacial shear stresses results in a diverging peeling force as $\theta \to 0$. This reflects the friction dissipation contribution proportional to $P^2 \cos^2\theta$ which leads to a much tougher peeling behavior at small peeling angle. In agreement with theoretical and experimental results [23], stiffer tapes entail higher peeling forces (Fig. 2a); nonetheless, $P$ value saturates at very large values of $\kappa$ as the Rivlin solution for rigid tapes [24] is approached. As expected, regardless of the specific interface behavior, the peeling resistance of the system increases with $\tilde{\gamma}$ (Fig. 2b), as higher energy is required to break the interfacial bonds. Interestingly, higher the value of $\tilde{\gamma}$, a more pronounced difference between the two models becomes evident for higher angles. This arises because elevated adhesion energies necessitate stronger forces for initiating detachment. Consequently, this heightened force extends the shear sliding zone, thereby augmenting the frictional work that needs to be surmounted during the peeling of the tape.

## 2.2 Effect of viscoelastic parameter and adhesion energy

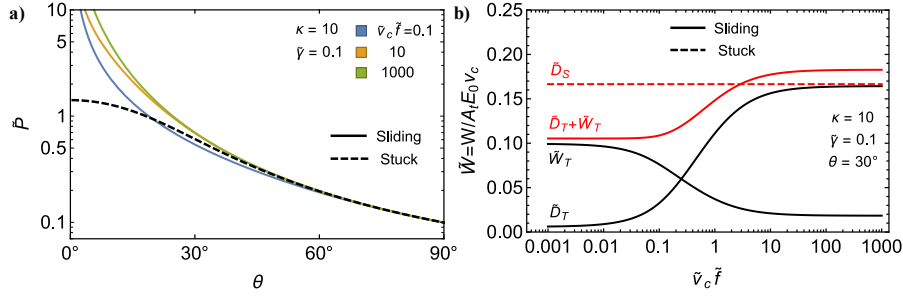

**Fig. 3.** (a) The dimensionless peeling force $\tilde{P}$ versus the peeling angle $\theta$ as a function of the dimensionless velocity parameter $\tilde{v}_c\tilde{f}$ predicted by stuck (dashed line) and siding (solid line) interface. (b) Dimensionless dissipative energy contribution for the stuck and sliding model as a function of $\tilde{v}_c\tilde{f}$.

As discussed so far, contrary to the stuck model, in the sliding approach the delamination resistance of the system explicitly depends on the peeling velocity and the frictional features of the interface. Fig. 3a shows $\tilde{P}$ vs $\theta$ for different values of the dimensionless velocity parameter defined as $\tilde{v}_c\tilde{f} = v_c f \tau / E_0 d$. Interestingly, for moderately high peeling angles $\theta$ and sufficiently small $\tilde{v}_c\tilde{f}$, despite the frictional losses, the sliding model predicts a lower value of the peeling force compared with the value observed in the stuck scenario (see $\tilde{v}_c\tilde{f} = 0.1$ curve). This is related to the different mechanisms of energy dissipation occurring in each case. Being $U_{el} = v_c A_t \sigma_0^2 / 2E_0$ the elastic energy per unit time stored in the tape for both the models, the viscoelastic dissipation per unit time can be evaluated as $D = W_I - U_{el}$. In what follow, $D_S$ and $D_T$ refer stuck and sliding model contributions, respectively. Fig. 3.4b shows the dimensionless dissipative energy terms as a function of the dimensionless velocity parameter for the stuck and sliding cases. For a stuck interface, the only source of energy dissipation arises from the viscoelastic creep occurring in the detached branch of tape, which is independent on $\theta$ and ($\tilde{D}_S$ curve in Fig. 3.4b). On the contrary, for a sliding interface, two additional sources of energy dissipation can be identified associated with the tape viscoelastic creep occurring in the sliding region, and the frictional dissipation. The first contribution is negligible for $\tilde{v}_c\tilde{f} \ll 1$, as in the sliding zone the tape is excited at a frequency which is much lower than the characteristic frequency of the material ($\omega \approx a/v_c \ll 1/\tau$), resulting in perfect elastic tape response with modulus $E_0$. Moreover, for sufficiently high peeling angles, the frictional losses $W_T$ has a low influence on the overall energy balance as they scale with $\cos^2\theta$. It results that, since in the sliding region the tape is pre-stretched, the viscoelastic creep in the detached portion occurs with a lower amount of deformation recovered compared with the stuck case and, in turn lower energy dissipation $\tilde{D}_T < \tilde{D}_S$.

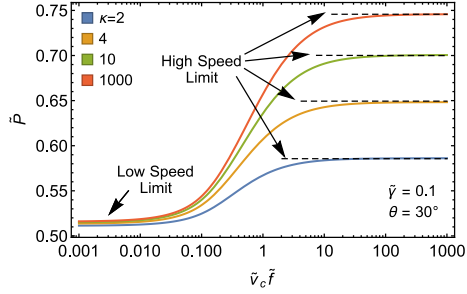

**Fig. 4.** (a) The dimensionless peeling force $\tilde{P}$ versus the dimensionless velocity parameter $\tilde{v}_c \tilde{f}$ predicted for different values of the viscoelastic parameter $\kappa = E_\infty/E_0$ in the case of sliding interface.

In Fig 4, the dimensionless peeling force $\tilde{P}$ as a function of the dimensionless velocity parameter $\tilde{v}_c \tilde{f}$ evaluated for fixed peeling angle $\theta = 30°$. Note that, the same trend of the quantity $\tilde{D}_T + \widetilde{W}_T$ (Fig. 3b) is observed. Such a behavior reflects the role of dissipative contributions in determining the overall peeling toughness of the system. Depending on the value of $\tilde{v}_c \tilde{f}$, three different regimes can be identified associated with a specific material behavior in the sliding zone. A low-speed ($\tilde{v}_c \tilde{f} \ll 1$) and an high-speed ($\tilde{v}_c \tilde{f} \gg 1$) plateaus are associated with a pure elastic response of the tape with $E_0$ and $E_\infty$ viscoelastic moduli, respectively. For intermediate values of $\tilde{v}_c \tilde{f}$, the hysteretic viscoelastic behavior governs the tape response, leading to the peeling force increasing with the velocity parameter following a power law $\tilde{P} \approx \left(\tilde{v}_c \tilde{f}\right)^n$, where $n$ is a function of the viscoelastic parameter $\kappa$. This result aligns with experimental evidences [25] where a similar velocity dependent peeling force is observed and, with empirical formulations based on a power law expression of the energy release rate [26,27].

## 3    Conclusions

We have investigated the single peeling behavior of a thin viscoelastic tape peeled away from a rigid substrate. In the first scenario, stuck contact condition has been assumed at the interface between tape and substrate. The overall viscoelastic peeling behavior is independent on the peeling velocity, with the peeling force resembling Kendall's prediction for elastic tape. In a second scenario the presence of frictional sliding at the interface close to the peeling front have been considered. The interplay between the tape hysteretic viscoelastic behavior and the frictional dissipations results in an enhancement of the peeling resistance at low peeling angles, and a velocity dependent peeling behavior, in agreement with phenomenological models and experimental observations.

21. B.-m. Zhang Newby, M. K. Chaudhury, Friction in adhesion, Langmuir 4 (17) (1998) 4865–4872.
22. N. Amouroux, J. Petit, L. L´eger, Role of interfacial resistance to shear stress on adhesive peel strength, Langmuir 17 (21) (2001) 6510–6517.
23. Lake GJ, Stevenson A. Wave phenomena in low angle peeling. J Adhes 1981;12(1):13–22.
24. Rivlin RS. The effective work of adhesion. In: Collected papers of RS Rivlin (2611–2614). New York, NY: Springer; 1997.
25. Krotova, Y. M. Kirillova, B. Deryagin, Issledovanie vliyaniya khimicheskogo stroeniya na adgeziyu, ZHURNAL FIZICHESKOI KHIMII 30 (9) (1956) 1921–1931.
26. M. Barquins, M. Ciccotti, On the kinetics of peeling of an adhesive tape under a constant imposed load, International journal of adhesion and adhesives 17 (1) (1997) 65–68.
27. G. Marin, C. Derail, Rheology and adherence of pressure-sensitive adhesives, The Journal of Adhesion 82 (5) (2006) 469–485.